# Stochastic Parameterization: Towards a new view of Weather and Climate Models

JUDITH BERNER[*]

*National Center for Atmospheric Research[,], Boulder, Colorado*

ULRICH ACHATZ

*Institut für Atmosphäre und Umwelt, Goethe-Universität, Frankfurt am Main, Germany*

LAURIANE BATTÉ

*CNRM-GAME, Météo-France/CNRS, Toulouse, France*

LISA BENGTSSON

*Swedish Meteorological and Hydrological Institute, Norrköping, Sweden*

ALVARO DE LA CÁMARA

*National Center for Atmospheric Research Boulder, Colorado*

HANNAH M. CHRISTENSEN

*University of Oxford, Atmospheric, Oceanic and Planetary Physics, Oxford*

MATTEO COLANGELI

*Gran Sasso Science Institute, Viale F. Crispi 7, 67100 L'Aquila, Italy*

DANIELLE R. B. COLEMAN

*National Center for Atmospheric Research[,], Boulder, Colorado*

DAAN CROMMELIN

*CWI Amsterdam, the Netherlands and Korteweg-de Vries Institute for Mathematics, University of Amsterdam*

STAMEN I. DOLAPTCHIEV

*Institut für Atmosphäre und Umwelt, Goethe-Universität, Frankfurt am Main, Germany*

CHRISTIAN L.E. FRANZKE

*Meteorological Institute and Centre for Earth System Research and Sustainability (CEN), University of Hamburg, Hamburg, Germany*

PETRA FRIEDERICHS

*Meteorological Institute, University of Bonn, Germany*

PETER IMKELLER

*Institut für Mathematik, Humboldt-Universität zu Berlin, Berlin, Germany*

HEIKKI JÄRVINEN

*University of Helsinki, Department of Physics, Helsinki, Finland*

STEPHAN JURICKE

*University of Oxford, Atmospheric, Oceanic and Planetary Physics, Oxford*

VASSILI KITSIOS

*CSIRO Oceans and Atmosphere Flagship, 107-121 Station St, Aspendale, Victoria 3195, AUSTRALIA*

FRANÇOIS LOTT

*Laboratoire de Météorologie Dynamique (CNRS/IPSL), Ecole Normale Supérieure, Paris, France*

VALERIO LUCARINI

*Meteorological Institute Centre for Earth System Research and Sustainability (CEN), University of Hamburg, Hamburg, Germany; Department of Mathematics and Statistics, University of Reading, Reading, UK*

SALIL MAHAJAN

*Oak Ridge National Laboratory, USA*

TIMOTHY N. PALMER

*University of Oxford, Atmospheric, Oceanic and Planetary Physics, Oxford*

CÉCILE PENLAND

*Physical Sciences Division, NOAA/Earth System Research Laboratory, Boulder, Colorado*

MIRJANA SAKRADZIJA

*Max Planck Institute for Meteorology and Hans-Ertel Centre for Weather Research, Deutscher Wetterdienst, Hamburg, Germany*

JIN-SONG VON STORCH

*Max-Planck Institute for Meteorology, Hamburg*

ANTJE WEISHEIMER

*National Centre for Atmospheric Science (NCAS), University of Oxford, Atmospheric, Oceanic and Planetary Physics, Oxford, and ECMWF, Shinfield Park, Reading, UK*

MICHAEL WENIGER

*Meteorological Institute, University of Bonn, Germany*

PAUL D. WILLIAMS

*Department of Meteorology, University of Reading, Reading, UK*

JUN-ICHI YANO

*GAME-CNRM, CNRS, Météo-France, 42 Av. Coriolis, Toulouse, France*
3

**ABSTRACT**

The last decade has seen the success of stochastic parameterizations in short-term, medium-range and seasonal forecasts: operational weather centers now routinely use stochastic parameterization schemes to better represent model inadequacy and improve the quantification of forecast uncertainty. Developed initially for numerical weather prediction, the inclusion of stochastic parameterizations not only provides better estimates of uncertainty, but it is also extremely promising for reducing longstanding climate biases and relevant for determining the climate response to external forcing.

This article highlights recent developments from different research groups which show that the stochastic representation of unresolved processes in the atmosphere, oceans, land surface and cryosphere of comprehensive weather and climate models (a) gives rise to more reliable probabilistic forecasts of weather and climate and (b) reduces systematic model bias.

We make a case that the use of mathematically stringent methods for the derivation of stochastic dynamic equations will lead to substantial improvements in our ability to accurately simulate weather and climate at all scales. Recent work in mathematics, statistical mechanics and turbulence is reviewed, its relevance for the climate problem demonstrated, and future research directions outlined.

**CAPSULE** (20-30 words)

Stochastic parameterizations - empirically derived, or based on rigorous mathematical and statistical concepts - have great potential to increase the predictive capability of next generation weather and climate models.

# 1 The need for stochastic parameterizations

Numerical weather and climate modeling is based on the discretization of the continuous equations of motion. Such models can be characterized in terms of their dynamical core, which describes the resolved scales of motion, and the physical parameterizations, which provide estimates of the grid-scale effect of processes, that cannot be resolved. This general approach has been hugely successful in that skillful predictions of weather and climate are now routinely made (e.g. Bauer et al. 2015). However, it has become apparent through the verification of these predictions that current state-of the art models still exhibit persistent and systematic shortcomings due to an inadequate representation of unresolved processes.

Despite the continuing increase of computing power, which allows numerical weather and climate prediction models to be run with ever higher resolution, the multi-scale nature of geophysical fluid dynamics implies that many important physical processes (e.g. tropical convection, gravity wave drag, micro-physical processes) are still not resolved. Parameterizations of sub grid-scale processes contain closure assumptions, and related parameters with inherent uncertainties. Although increasing model resolution gradually pushes these assumptions further down the spectrum of motions, it is realistic to assume that some form of closure or physical parameterization will be present in simulation models into the foreseeable future.

Moreover, for climate simulations, a decision must be made as to whether computational resources should be used to increase the representation of subgrid physical processes or to build a comprehensive Earth-System Model, by including additional climate components such as the cryosphere, chemistry and biosphere. In addition, the decision must be made about whether computational resources should go towards increased horizontal, vertical and temporal resolution or additional ensemble members.

Additional challenges are posed by intrinsically coupled phenomena like the Madden-Julian Oscillation (MJO) and tropical cyclones. Correctly simulating these tropical multi-scale features requires resolving or accurately representing small-scale processes such as convection in addition to capturing the large-scale response and feedback. Many of the Coupled Model Intercomparison Project phase 5 (CMIP5) climate models still do not properly simulate the MJO and convectively coupled waves (Hung et al. 2013).

A great challenge is posed by the representation of partially resolved processes (either in the time or space domain). The range of scales on which a physical process is only partially resolved is often referred to as the "gray zone" (e.g. Gerard 2007). In this gray zone, the number of eddies in each grid box is no longer large enough to fulfill the "law of large numbers" underlying deterministic bulk parameterizations and a stochastic approach becomes essential. An example for a partially resolved process is convection, which is often split into a resolved (large-scale) and parameterized component (e.g. Arakawa 2004). The equilibrium assumption no longer holds when the model resolution is increased such that a clear scale separation between convection and larger scales no longer is valid (e.g.Yano and Plant 2012a,b). In this case the subgrid-scale parameterization takes a prognostic form rather than being diagnostic, as explicitly shown for the mass-flux formulation by Yano (2014).

As the next generation of numerical models attempts to seamlessly predict weather as well as climate, there is an increasing need to develop parameterizations that adapt automatically to different spatial scales ("scale-aware parameterizations"). A big advantage of the mathematically rigorous approach is that the subgrid-model is valid for increasing spatial resolutions within a range of scales that is obtained as part of the derivation.

Mathematical approaches to stochastic modeling rely on the assumption that a physical system can be expressed in terms of variables of interest, and variables which one does not want to explicitly resolve. In the mathematical literature this is usually referred to as the operation of coarse-graining and performed through the method of homogenization (Papanicolaou and Kohler 1974, Gardiner 1985, Pavliotis and Stuart 2008). The goal is then to derive an effective equation for the slow predictable processes and to represent the effect of the now unresolved variables as random noise terms.

Many stochastic parameterizations are based on the assumption of a scale separation between the temporal decorrelation rates between the rapidly fluctuating processes represented by a white noise and the slow processes of interest (e.g., Gardiner, 1985; Penland, 2003a). In geophysical applications, there is often - but not always – a relationship between spatial and temporal scales of variability, with fast processes associated with small scales and slow processes associated with large scales. If this is the case, separating physical processes by timescales can result in decomposing small-scale features from large-scale phenomena and spatial and temporal scale separation become equivalent.

Such a thinking underlies the pioneering study of Hasselmann (1976), who split the coupled ocean-atmosphere system into a slow ocean and fast weather fluctuation components and subsequently derived an effective equation for the ocean circulation only. One finds that the impact of the fast variables on the dynamics of the slow variables boils down to a deterministic correction – a mean field effect sometimes referred to as noise-induced drift or rectification – plus a stochastic component, which is a white random noise in the limit of infinite time scale separation.

A simple example demonstrating noise-induced transitions and drifts is presented in Figure 1. Assume, the unforced nonlinear climate system can be described by a double-well potential (a). If the noise is sufficiently small (denoted by short red arrows) and under appropriate initial conditions, the system will stay for a finite time in the deeper potential well and the associated probability density function of states will have a single maximum (b). As the amplitude of the noise increases (long arrows in c), the system can undergo a noise-induced transition and reach the secondary potential well. The resulting probability density function (PDF) will exhibit two local maxima (d), signifying two different climate regimes, rather than a single maximum, as in the small-noise scenario. Note, that the stochastic forcing not only changes the variance, but also the mean.

But even a linear system characterized by a single potential when unforced can change the mean, if forced by multiplicative or state-dependent white noise (e-h). The noise is called "multiplicative", if its amplitude is a function of the state, which is denoted by the red errors of different length in panel g. The noise-induced drift changes the single-well potential of the unforced system (e), so that the effective potential including the effects of the multiplicative noise has multiple wells (not shown) and the associated PDF becomes bimodal (h). Note, that in this example the shift in the mean compared to the unforced PDF (f) is caused by the noise, which is referred to as "noise-induced drift" (see e.g., Sardeshmukh et al. 2001, Berner 2005, Sura et al. 2005).

Operational weather and climate centers use now stochastic parameterization schemes routinely to make ensemble predictions from short-range to seasonal time scales (e.g., Berner et al. 2009, Weisheimer et al. 2014). Most ensembles suffer from underdispersion, which means that - on average – the observed state is more often outside the cone of forecasts than can be statistically justified. Stochastic perturbations introduce more diversity among the forecasts, which helps to ameliorate this problem and result in more skillful ensemble forecasts.

A fundamental argument, that has been often overlooked, is that the merit of stochastic parameterization goes far beyond providing uncertainty estimations for weather and climate predictions, but may be also needed for better representing the mean state (e.g., Sardeshmukh et al. 2001, Palmer 2001, Berner et al. 2008) and regime transitions (e.g., Williams et al. 2003, 2004, Birner and Williams 2008, Christensen et al. 2015a) via inherent non-linear processes. This is especially relevant for climate projections, which have long-standing mean state errors, such as e.g., a double inter-tropical convergence zone (e.g., Lin 2007), and erroneous stratocumulus cloud cover, which play a crucial role in the climate response to external forcing.

Mechanisms for how Gaussian zero-mean fluctuations can change the mean state (see Figure 1) have been discussed e.g. in Tompkins and Berner (2008) and Beena and von Storch (2009). Tompkins and Berner (2008) introduce perturbations to the humidity field and find that positive perturbations are more likely to trigger a convective event than negative perturbations can suppress convection. Beena and von Storch (2009) study the ocean response air-sea flux perturbations and similarly find that negative buoyancy anomalies result in an altered stratification, while positive anomalies tend to sustain the existing stratification. Insofar as stochastic parameterizations can change the mean state, they have the potential to affect the response to changes in the external forcing (e.g., Seiffert and von Storch 2008).

In mathematical terms, this is the question how a stochastic forcing affects the invariant measure of a deterministic dynamical system (Lucarini 2012) and how the climate response to such a forcing can be framed as a problem of non-equilibrium statistical mechanics (Colangeli et al. 2012, 2014, Lucarini and Sarno 2011, Lucarini et 2014a,b).

Here, we argue, that stochastic parameterizations are essential for:

• Estimating uncertainty in weather and climate predictions

• Reducing systematic model errors arising from unrepresented subgrid-scale fluctuations

• Triggering noise-induced regime transitions

• Capturing the response to changes in the external forcing

and should be applied in a systematic and consistent fashion, not only to weather, but also to climate simulations.

Several studies have identified the assessment of the benefits of stochastic closure schemes as key outstanding challenge in the area of mathematics applied to the climate system (Palmer 2001, 2012, Palmer and Williams 2008, Williams et al. 2013). For accessible reviews of rigorous mathematical approaches applied to weather and climate, we refer to Penland (2003a,b), Majda et al. (2008) and Franzke et al. (2015). The current study focuses on recent developments and successful applications of empirical and rigorous approaches to the subgrid-parameterization problem in weather and climate models.

## 2 Representing Uncertainty in Comprehensive Climate and Weather Models

*2.1 Adding uncertainty a posteriori: the stochastically perturbed parameterization tendency scheme and the stochastic kinetic-energy backscatter scheme*

Stochastic parameterizations are based on the notion that – as spatial resolution increases – the method of averaging (Arnold 2001, Monahan and Culina 2011) is no longer valid and the subgrid-scale variability should be sampled rather than represented by the equilibrium mean. In addition, unrepresented interactions between unresolved subgrid-scale processes with the large-scale flow might affect the resolved dynamics.

Former is addressed by the stochastically perturbed parameterization tendency (SPPT) scheme, which perturbs the net tendencies of the physical process parameterizations (convection, radiation, cloud physics, turbulence and gravity wave drag). One essential feature for its success is that the noise is correlated in space and time. SPPT has a beneficial impact on medium range, seasonal and climate forecasts (Buizza et al. 1999, Teixeira and Reynolds 2008, Palmer et al. 2009, Weisheimer et al. 2014, Christensen et al. 2015b, Dawson and Palmer 2015, Batté and Doblas-Reyes 2015). SPPT tends to be most active in the tropics and near the surface, where the parameterized tendencies are large.

The stochastic kinetic-energy backscatter scheme (SKEBS) aims to represent model uncertainty arising from unresolved subgrid-scale processes and their interactions with larger scales by introducing random perturbations to the streamfunction and potential temperature tendencies. For this purpose, the scheme re-injects a small fraction of the dissipated energy into the resolved flow. Originally developed in the context of Large Eddy Simulations (LES; Mason and Thomson 1992), it was adapted by Shutts (2005) for Numerical Weather Prediction (NWP).

Depending on the details of the implementations, SKEBS tends to have most impact in the storm-track regions and in the free atmosphere above the boundary layer and permits the physical parameterization schemes to adjust to a slightly perturbed large-scale background flow. Its beneficial impact on weather and climate forecasts are reported e.g., in Berner et al. (2011, 2015), Tennant et al. (2011), Weisheimer et al. (2014), Sanchez et al. (2015); albeit Shutts (2013) criticizes the arbitrary nature of some of the design features based on coarse-graining high-resolution simulations to compute the backscatter term. His stochastic convective backscatter scheme (Shutts, 2015) includes a phase relationship between flow and perturbations and adds additional perturbations to the divergent flow to remedy some of the identified shortcomings.

While these schemes are motivated by physical reasoning and scheme parameters are informed in some manner, for example by coarse-graining high-resolution output (Shutts and Palmer 2007, Shutts and Callado Pallarès 2014) or comparison with observations (Watson et al. 2015), the perturbations are essentially empirical constructs. For example, the amplitude of the perturbations is typically determined as the value that satisfactorily reduces the ensemble underdispersion. Obviously such an approach is only possible for forecast ranges where verification is possible, such as for short-term, medium-range and seasonal forecasts. A common criticism of this approach is that the improved skill is solely the result of the increase in spread. However, Berner et al. (2015) found that the merits of stochastic parameterization go beyond increasing spread and can account for structural model uncertainty.

In the following examples, we show recent results that demonstrate the potential of stochastic parameterizations to improve the mean state representation and variability as well as the skill of seasonal forecasts.

First, we present recent results from the seasonal forecasting system at ECMWF. In the simulations with both, SPPT and SKEBS, excessively strong convective activity over the Maritime Continent and the tropical Western Pacific is reduced, leading to smaller biases in outgoing longwave radiation (Figure 2, adapted from Weisheimer et al. 2014), cloud cover, precipitation and near-surface winds. The stochastic schemes also lead to an increase in the frequency (Figure 3, from Weisheimer et al. 2014) and amplitude of MJO events, which is an improvement. A reduction of excessive amplitudes in westward propagating convectively coupled waves in an earlier model version is reported in Berner et al. 2012.

Another example of the positive impact of stochastic schemes is evident in climate simulations with the Community Earth System Model (CESM). Compared to observations, the modeled spectrum of average sea surface temperature in the Nino 3.4 region has three times more power for periods between 2 and 4 years (Figure 4, adapted from Christensen et al. 2016). SPPT markedly reduces the temperature variability in this frequency range, leading to a much better agreement with nature (Christensen et al., 2016). Interestingly, in these examples adding stochasticity results in *reduced* variability, which is a non-trivial response.

Along with the improved model climate, stochastic perturbations benefit probabilistic forecast performance on seasonal timescales. This has been reported in a number of studies using earlier versions of ECMWF's seasonal system (Berner at al. 2008, Dobles-Reyes et al. 2009, Palmer at al. 2009) and recently been confirmed in the newest version (Weisheimer et al. 2014) and in the EC-Earth system model (Batté and Doblas-Reyes 2015). Figure 5 shows ensemble mean and spread in forecasts for Nino 3.4 area sea-surface temperatures with the EC-Earth model, run at a standard horizontal resolution (SR, ca. 60km for the atmospheric and ca. 100km for the ocean component) and a high resolution (HR, ca. 40km for the atmospheric component and 25km for the ocean.) For both resolutions, the introduction of SPPT perturbations increases the ensemble spread. Furthermore, SPPT reduces the mean error in the standard resolution, but not as much as increasing horizontal resolution.

A number of studies have found evidence for stochasticity leading to noise-induced transitions in mid-latitude circulation regimes, especially over the Pacific-North America region (Jung et al. 2005, Berner et al. 2012, Dawson and Palmer, 2015, Weisheimer et al. 2014). These results suggest that stochastic parameterizations are also relevant for the prediction of the dominant modes of atmospheric variability such as the North Atlantic Oscillation and the Pacific North American pattern (Berner, unpublished mansucript).

*2.2 Adding uncertainty a priori: perturbed parameter approaches for the atmospheric component*

While the performance of the stochastic schemes discussed in the last section is undisputed, they have been criticized in that they are added *a posteriori* to models that have been independently developed and tuned. Ideally, stochastic perturbations should represent model uncertainty where it occurs. One obvious way to represent uncertainty at its source rather than *a posteriori* is the perturbed parameter approach, which perturbs the closure parameters in the physical process parameterizations. There are two variants: the parameter can be fixed throughout the integration, but vary for each ensemble member (e.g. Murphy et al. 2004, Hacker et al. 2011a) or vary randomly with time (e.g. Bowler et al. 2008, 2009, Ollinaho et al. 2013, Jankov et al., 2016). Strictly, the first variant is not a stochastic parameterization, but an example for a multi-model, since each ensemble member has a different climatology. However, since stochastic parameter perturbations are routinely compared to fixed-parameter schemes, this section discusses both.

While perturbed-parameter ensembles typically outperform unperturbed ensembles on weather timescales, they typically cannot sufficiently account for all deficiencies in the spread (Hacker et al. 2011, Reynolds et al. 2011, Christensen et al. 2015b) and do not lead to the same reliability as the *a posteriori* schemes discussed above (Berner et al., 2015). Presumably, this is due to the fact that *a posteriori* schemes are designed to encapsulate all model uncertainty, of which parameter uncertainty is only one contributor.

An ensemble system is considered statistically reliable when a predicted probability for a particular event (e.g. temperature exceeding 17°C) compares well with the observed frequencies. Another limitation of this approach is that the parameter uncertainty estimates are subjective, and information about parameter interdependencies is not included.

The following studies are examples for applications of the perturbed-parameter approach to physical process parameterizations and perturbing the interface between different model components. We start with results pertaining to perturbations in the atmospheric component and move to those of other model components, such as land and ocean models, which are more relevant for climate applications.

A number of studies report on improved skill due to parameter perturbations to boundary layer and convection schemes (Hacker et al. 2011, Reynolds et al. 2011, Jankov et al., 2016). Recently, a stochastic "eddy-diffusivity/ mass-flux" parameterization has been developed (Suselj et al. 2013, 2014), which combines an eddy-diffusivity component with a stochastic mass-flux scheme. The resulting scheme unifies boundary layer and shallow convection and was operationally implemented in the operational Navy Global Environmental Model.

Christensen et al. (2015b) used an objective covariance estimate of parameter uncertainty (Järvinen et al. 2012, Ollinaho et al. 2013) for four convection closure parameters and developed both a fixed-parameter and a stochastically varying perturbation scheme. Both schemes improved the forecast skill of the ECMWF ensemble prediction system, with a larger impact observed for the fixed perturbed parameter scheme (Figure 6, adapted from Christensen et al. 2015b). In addition, for some variables such as wind at 850hPa, the scheme leads to a reduction in bias (Figure 6, adapted from Christensen et al. 2015b).

Recently, a body of work proposes stochastic approaches for another atmospheric parameterization, namely non-orographic gravity waves (Lott et al. 2012, Lott and Guez 2013, and de la Cámara and Lott 2015). Observational studies indicate that the gravity wave field is very intermittent and only predictable in a statistical sense. Recently, de la Cámara et al. (2014) informed the free parameters of the stochastic gravity-wave scheme using momentum flux measurements.

*2.3 Uncertainty in land surface, ocean and coupled component models*

Physical parameters of land surface models are often not well constrained by observations. A recent study by MacLeod et al. (2015) introduced parameter perturbations to key soil parameters, and compared their impact with stochastic perturbations of the soil moisture tendencies in seasonal forecasts with the ECMWF coupled model. Both the perturbed parameter approach and the stochastic tendency perturbations improved the forecasts of extreme air temperature for the European heat wave of 2003.

A shortcoming in land models stems from the omission of sub-grid land heterogeneity, which impacts the surface heat flux. Langan et al. (2014) retained the subgrid-variability by drawing the area for each plant functional types at each timestep from a Dirichlet distribution, rather than using constant area weights. First results with a single column model version of CESM reveal an increase in the variability as well as larger extreme values in convective precipitation (Figure 7, adapted from by Langan et al. 2014).

The coupled atmosphere-ocean system is very sensitive to fluctuations in the fluxes between its component models. Air-sea fluxes of buoyancy, energy, and momentum vary on a vast range of space and time scales, including scales that are too small or fast to be explicitly resolved by global climate models. For example, convective clouds in the atmosphere will cause subgrid fluctuations at the air-sea interface, in both, the downward fresh water flux and short-wave solar radiation. The response of the climate to stochastic perturbations of the air-sea buoyancy flux is studied by Williams (2012) in a coupled atmosphere-ocean model. The response is complex and involves changes to the oceanic mixed-layer depth, sea-surface temperature, atmospheric Hadley circulation, and fresh water flux across the sea surface (Figure 8, from Williams 2012). These findings suggest that the lack of representation of stochastic subgrid variability in air-sea fluxes may contribute to some of the biases exhibited by contemporary coupled climate models.

Since the buoyancy effects in the ocean are different from that in the atmosphere, the length scale at which rotational effects become as important as gravity wave effects is much smaller. Consequently, mesoscale eddies in state-of-the art ocean models are still far from being resolved and are usually represented by traditional bulk parameterizations (Gent and McWilliams 1990, Redi 1982). A recent study by Li and von Storch (2013) computes the contributions from the mean and fluctuating component of heat flux divergence in a high-resolution ocean model. The magnitude of the fluctuations is about one order of magnitude larger than the mean component (Figure 9, adapted from Li and von Storch 2013) suggesting that classical parameterizations significantly underestimate the total eddy flux. The fluctuating part, even though having zero mean, can play an important role in generating large-scale low-frequency variations and in shaping the mean oceanic circulation.

Juricke et al. (2013) and Juricke and Jung (2014) recently investigated the sensitivity of an ocean-sea ice model to variations in the ice strength parameter. As this parameter is not observable, large uncertainties remain in the choice of its value, although it is very important for modeling sea ice drift. Varying this parameter stochastically results in changes to the mean sea ice distribution as well as sea ice spread. Compared to perturbations of the atmospheric initial conditions, the incorporation of additional stochastic ice strength perturbations leads to a considerably increase in spread of the simulated sea ice thickness in the central Arctic (Figure 10, adapted from Juricke et al. 2014), which is a better match with the observed uncertainties (Juricke et al. 2014).

*2.4 Data Assimilation and Extreme Events*

The purpose of data assimilation is to combine observations with short-term model-forecasts to come up with a gridded and physically consistent estimate of the state of the atmosphere, also called "analysis". One method is to use short-term forecasts as the first guess fields in ensemble data assimilation. As such, ensemble data assimilation inherits the shortcomings of short-term ensemble predictions, namely, the underdispersivness in the spread. Recent work has demonstrated that the stochastic parameterizations that are beneficial for ensemble prediction, can also improve analyses fields (Isaksen et al. 2007, Houtekamer et al. 2009, Mitchell and Gottwald 2012, Hamill and Whitaker 2011, Ha et al. 2015, Romine at al. 2015). In particular, Ha et al. 2015 showed that the inclusion of a stochastic parameterization improved the mean analysis, even if the observations were constrained to those in the control experiment. A cutting-edge frontier is the use of memory effects in Kalman filter data assimilation schemes (O'Kane and Frederiksen 2012).

The impact of stochastic perturbations on extremes has only been considered very recently. A body of work focuses on the description of non-Gaussian subgrid-scale processes (Majda et al. 2009, Sardeshmukh and Sura 2009, Sura 2011, Sardeshmukh et al. 2015). Franzke (2012) showed that his reduced stochastic model (see next section) captures the extremes of the full model. He et al. (2012) studied the influence of an explicitly stochastic representation of mixing in the stable boundary layer on the extremes of near-surface wind speed in a single column model. Tagle et al. (2015) were the first to study the effect of the stochastic parameterizations in a comprehensive climate model. They found that the stochastic parameterizations had a big impact on the surface temperature mean and variability, but hardly changed the tail behavior.

## 3 Systematic mathematical and statistical physics approaches

This section introduces systematic mathematical and statistical physics approaches to the parameterization problem and reports on recent work on the application of these rigorous methods to the weather and climate system.

### *3.1. Mathematical and Numerical implications of stochasticity*

Although the motions of the atmosphere and ocean are described by the Navier-Stokes equation, large-scale flows can often be modeled under hydrostatic approximation. This leads to the deterministic primitive equation system. If we want to represent continuous small-scale fluctuations as stochastic terms, these equations need to be generalized to allow for stochasticity. A relevant mathematical field is thus the extension of the derivation to the stochastic primitive equations for two-dimensional (Ewald et al. 2007; Glatt-Holtz and Ziane 2008; Glatt-Holtz and Temam 2011) and three-dimensional flows (Debussche et al. 2012).

Moreover, stochastic systems require calculi and numerical schemes fundamentally different from the ones available to solve deterministic systems. The two most commonly used stochastic integral types are the Itô-integral (Itô 1951) and the Stratonovich-integral (Stratonovich 1966). When the fast processes of a continuous system are modeled by white noise – as is common for physical applications - the resulting stochastic model converges to a Stratonovitch stochastic differential equation (Wong and Zakai 1965, Papanicolaou and Kohler 1974, Gardiner 1985, Penland 2003a,b). Discrete systems converge to the Itô stochastic differential equation. Starting in the 1970s a solid framework of numerical methods for stochastic ordinary differential equations was developed (Rümelin 1982, Kloeden and Platen 1992, Milstein 1995, Kloeden 2002). However, this has been extended to high-order schemes only recently (Jentzen and Kloeden 2009, Weniger 2014). With stochastic parameterizations becoming more common in weather and climate simulations, a revision of the deterministic numerical schemes should be undertaken to ensure the convergence of the numerical solutions.

*3.2 Homogenization and stochastic mode reduction*

Numerical weather and climate modeling can be seen as a model reduction problem. Because we cannot numerically solve the full continuous equations, we have to truncate the equations at some scale and then treat the unresolved processes in some smart way. A systematic approach for the derivation of reduced order models from first principles is performed through the method of homogenization or adiabatic elimination (Wong and Zakai 1965, Khas'minskii 1966, Kurtz 1973, Papanicolaou and Kohler 1974, Pavliotis and Stuart 2008). The fundamental idea is to decompose the state vector into slow and fast components, represent the fast processes by a stochastic term and derive analytically an effective equation for the slow, predictable modes.

Majda et al. (1999) and Majda et al., 2001 expanded this body of work by making additional assumptions on the nonlinear self-interaction of the fast modes and coined the term "stochastic mode reduction".

The stochastic mode reduction has been demonstrated to successfully model regime-behavior and low-frequency variability for conceptual models of the atmosphere (Majda et al. 2003), the barotropic vorticity (Franzke et al. 2005) and a quasi-geostrophic three-layer model on the sphere with realistic orography (Franzke and Majda 2006). However, due to both, the sheer amount of analytical derivation and the compute-memory requirement in the numerical implementation of the resulting equations, the stochastic mode reduction cannot be easily applied to comprehensive climate models of arbitrary complexity. A possible way forward is to apply the stochastic mode reduction locally at each gridpoint rather than globally (Dolaptchiev et al. (2013 a,b).

These mathematical techniques are rigorously valid only in the limit of large time-scale separation, although some studies report good empirical results, even when this condition is not, or only partly met (Dozier and Tappert 1978a,b , Majda et al. 2003 2008, Franzke et al. 2005, Franzke and Majda 2006). When the time scale separation between the fast and slow processes is not too large, the picture of the parameterization as being constructed as the sum of a suitably defined deterministic plus random corrections has to be amended to take memory effects into account (e.g. Zwanzig 2001, Chekroun et al. 2015a,b). Unfortunately, the condition of scale separation is typically not met in geophysical fluid dynamics applications (Sardeshmukh and Penland 2015, Yano 2015, Yano et al. 2015), which poses limitations to the application of homogenization. An alternative, which does not make any assumptions about time scale separation and provides an explicit expression for the terms responsible for the memory effect is proposed by Wouters and Lucarini (2012, 2013), who, instead, assume the presence of a weak dynamical coupling between the fast and the slow scales of motion.

The question of which stochastic process is best suited to describe the nonlinear interactions of the unresolved processes is an open topic. While methods for Gaussian diffusion processes are well known (Oppenheim 1975) it may be the case that other formulations like Lévy processes are better suited to describe the underlying physics. For the interested reader, we refer to recent studies by Penland and Ewald 2008, Penland and Sardeshmukh 2012, Hein et al. 2010, Gairing and Imkeller 2012, 2013, Thompson et al. 2015.

*3.3 Adaptation of Concepts from Statistical Physics to Weather and Climate*

The scale-aware representation of convection and clouds on high-resolution grids (1-50 km) has been a long-standing challenge for weather and climate models. Within a single model column, convection is not uniquely determined by the resolved-scale processes, and the distribution of possible realizations of subgrid-scale convection highly depends on model resolution. Furthermore, horizontal transports of heat, moisture or momentum from neighboring grid-boxes are typically neglected. Thus, to achieve scale-awareness, it is necessary to represent scale-dependent convective fluctuations about the ensemble average response. In addition, because of the lack of time-scale separation, a correct representation of convection across scales requires memory of subgrid-states from previous time steps.

A novel approach to represent the fluctuations in an ensemble of deep convective clouds adapts concepts from statistical mechanics (Craig and Cohen 2006). Based on this theory, a stochastic parameterization of deep convection was developed to represent fluctuations of the subgrid convective mass flux about statistical equilibrium (Plant and Craig 2008). This is especially attractive for variable-resolution grids, since the statistics automatically adapt to the grid-resolution. This approach was extended to shallow convective clouds by introducing a memory effect arising from the correlation between the cloud mass fluxes and cloud lifetimes (Sakradzija et al. 2015). Figure 11 (adapted from Sakradzija et al. 2015) shows histograms of the subgrid cloud-base mass flux in the stochastic shallow cumulus cloud scheme and coarse-grained large-eddy simulation at different horizontal resolutions. The histograms match closely and are scale-aware.

*3.4 Modeling convective processes by Markov chains and cellular automata*

Another way to introduce temporal memory and nonlocal effects is the use of Markov chains and cellular automata. A Markov chain is a mathematical system that undergoes transitions from one *discrete* state to another and the probabilities associated with the various state changes are called transition probabilities. If observational data or high-resolutions simulations are used to inform the transition probabilities, the Markov chains are called data-driven.

An example for this approach is the "stochastic convective parameterization" which describes the convective state of the entire model column as a discrete Markov chain. (Khouider et al. 2010, Dorrestijn et al. 2013a,b, 2015, Gottwald et al. 2015). The system can only reside in a few distinct convective states – e.g., but not necessarily: clear sky, shallow or deep convection, - and the random transitions from one state to another evolve as a Markov chain. For example, Dorrestijn et al. 2013a cover the horizontal domain of the numerical model with a high-resolution lattice (with typical lattice spacing of 100m to 1000m), and on each lattice node lives a copy of the discrete stochastic process for the convective state (Figure 12, adapted from Dorrestijn et al. 2013a). The transition probabilities are estimated from a cloud-resolving LES model. By averaging over blocks of lattice nodes, convective area fractions and related quantities can be obtained for spatial domains of arbitrary size. The resulting patterns and temporal behavior of the area fractions are quite realistic. Furthermore, the formulation on a high-resolution lattice (or microlattice) makes it possible to compute convective fractions for varying area sizes, so that a parameterization based on these fractions is scale-adaptive.

Frenkel et al. (2012) and Peters et al. (2013) use the stochastic model introduced in Khouider et al. (2010), but different methods to estimate the transition probabilities. Khouider et al (2010) and Frenkel et al. (2012) formulate the rules based on physical insight, while Peters et al. (2013) use observations for their estimates. Latter find that the estimates from observation can notably differ from those based on physical intuition.

A related approach are cellular automata which are often used as simple mathematical models to simulate spatial self-organizational behavior such as convective organization A cellular automaton describes the evolution of discrete states on a lattice grid. The states are updated according to a set of rules based on the states of neighboring cells at the previous time step. In addition to memory, cellular automata can allow for lateral communications between neighboring grid boxes and thus introduce spatial correlations.

The idea of using cellular automata within NWP was first proposed by Palmer (2001) and first applications used them as a quasi-stochastic pattern generator for SKEBS (Shutts 2005, Berner et al. 2008). Bengtsson et al. (2013) pioneered the use of a cellular automaton for the parameterization of convection, which allows for the horizontal transports of heat, moisture and momentum across neighboring grid-boxes. The scheme has been shown to enhance the organization of convective squall-lines (Bengtsson et al. 2013) and improves the skill of accumulated precipitation in a high-resolution ensemble prediction system (Bengtsson and Körnich 2015).

*3.5 Climate Response in the presence of small-scale fluctuations*

While there is extensive work focusing on the response of the climate system to changes in the external forcing, either natural - such as the forcing from a localized tropical heating as it occurs in Nino - or anthropogenic - such from increased greenhouse gases, little attention has been given to the fact if and how the representation of the subgrid-scale can alter that response. In the mathematical community, this is the topic of response theory and the fluctuation-dissipation theorem (e.g., Marconi et al. 2008, Lacorata and Vulpiani 2007, Colangeli et al. 2011, Lucarini and Colangeli 2012, Colangeli and Lucarini 2014).

Seiffert and von Storch (2008) were the first to investigate the response of a climate model to CO2-forcing in the presence of subgrid-scale fluctuations in atmospheric temperature, divergence and vorticity. In their model, the strength of the global warming due to a CO2-doubling is altered by up to 15% near the surface and up to 25% in the upper troposphere (Figure 13, from Seiffert and von Storch 2008) depending on the exact representation of the small-scale fluctuations. Applying a stochastic model to their simulations, they found that the small-scale fluctuations change the temperature response via a statistical damping that acts as a restoring force. In addition, the small-scale fluctuations can affect feedback and interaction processes that are directly coupled to an increase in CO2, thereby altering the CO2-related radiative forcing (Seiffert and von Storch 2010).

The fluctuation-dissipation theorem (FDT) is concerned with the response of a system to small changes in the forcing. In particular, it tries to relate the response to the natural fluctuations in the system (Kubo 1966, Deker and Haake 1975, Hänggi and Thomas 1977, Leith 1975, Risken 1984). In the atmospheric sciences, the FDT-operator is estimated from model output, in particular the variances and covariances of the state variables at different time lags. The so obtained empirical linear model is able to predict the response to changes in the external forcing, such as signature from localized tropical heat forcing (Gritsun and Branstator 2007, Gritsun et al. 2008).

Achatz et al. (2013) argue that subgrid-scale parameterizations developed for a present day climate, might no longer be accurate in a changing climate. They use the FDT to adjust the subgrid-scale representation of the forced system. Figure 14 (adapted from Achatz et al. 2013) shows that a low-order model with a subgrid-scale parameterization corrected by the FDT yields a better response in streamfunction variance than without the correction.

While some success of FDT-techniques to low-frequency climate modeling has been demonstrated, some of the mathematical assumptions are not strictly met. Recent work expands the mathematical underpinning by formulating the response theory more generally and is better suited for non-equilibrium systems (Ruelle 2009, Lucarini and Sarno 2011) and climate projections (Lucarini et al. 2014b, Ragone et al. 2015).

*3.6 Statistical Dynamical Closure Theory*

Kraichnan (1959) first illustrated that renormalization of the statistical equations of fluid motion can been used to produce self-consistent parameterizations of the subgrid turbulent processes. It is on this basis that Frederiksen and Davies (1997) developed stochastic parameterisations of subgrid turbulence in barotropic atmospheric simulations on the sphere. The subgrid parameterizations consist of drain, backscatter and net eddy viscosities, which are determined from the statistics of higher resolution closure simulations. The aim here is that the spectra from the low-resolution simulation with stochastic subgrid parameterization should be ideally statistically indistinguishable from those produced by the high-resolution simulation, which would allow to save computational resources. Implementation of this approach into an atmospheric GCM resulted in significantly improved circulation and energy spectra (Frederiksen et al. 2003). These ideas were further developed by Frederiksen (1999, 2012a,b), and O'Kane and Frederiksen (2008).

Frederiksen and Kepert (2006) then used the functional form of these closure approaches to develop a zero-parameter stochastic modeling framework, where the eddy viscosities are determined from higher-resolution reference simulations. This approach was successfully applied to baroclinic geophysical simulations in Zidikheri and Frederiksen (2009, 2010a,b). Recently, Kitsios et al. (2012, 2013,2014) used this approach to determine the eddy viscosities from a series of high-resolution atmospheric and oceanic reference simulations. The isotropized versions of the subgrid-eddy viscosities were then characterized by a set of scaling laws. Large eddy simulations with subgrid models defined by these scaling laws (solid lines in Figure 15) were able to reproduce the statistics of the high-resolution reference simulations (dashed lines in Figure 15) across all resolved scales. This demonstrates that including a stochastic subgrid parameterization in the low-resolution simulations makes them indistinguishable from the high-resolution reference.
The scaling laws further enable the subgrid parameterizations to be utilized more widely, as they remove the need to generate the subgrid coefficients from a reference simulation.

**Concluding Remarks**

In this article, we attempt to narrow the gap between the fields of numerical meteorological models and applied mathematics in the development of stochastic parameterizations: on the one hand geo-scientists are often unaware of mathematically rigorous results that can aid in the development of physically relevant parameterizations, on the other hand mathematicians often do not know about open issues in scientific applications that might be mathematically tractable.

Over the last decade or two, increasing evidence has pointed to the potential of this approach, albeit applied in an *ad hoc* manner and tuned to specific applications. This is apparent in the choices made at operational weather centers, where stochastic parameterization schemes are now routinely used to represent model inadequacy better and improve probabilistic forecast skill. Here, we revisit recent work that demonstrates that stochastic parameterizations are not only essential for the estimation of the uncertainty in weather forecasts, but are also necessary for accurate climate and climate change projections. Stochastic parameterizations have the potential to reduce systematic model errors, trigger noise-induced regime transitions, and modify the response to changes in the external forcing.

Ideally, stochastic parameterizations should be developed alongside the physical parameterization and dynamical core development and not tuned to yield a particular model performance. This approach is hampered by the fact that parameters in climate and weather are typically adjusted ("tuned") to yield the best mean state and/or the best variability. This can result in compensating model errors, which pose a big challenge to model development in general, and stochastic parameterizations in particular. A stochastic parameterization might improve the model from a process perspective, but its decreased systematic error no longer compensates other model errors, resulting in an overall larger bias (Palmer and Weisheimer 2011, Berner et al. 2012). Clearly, such structural uncertainties need to be addressed in order to improve the predictive skills of our models.

Mathematically rigorous approaches decompose the system-at-hand into slow and fast components. They focus on the accurate simulation of the large, predictable scales, while only the statistical properties of the small, unpredictable scales need to be captured. One finds that the impact of the fast variables on the dynamics of the slow variables boils down to a deterministic correction plus a stochastic component. This immediately points to the fact that the classical parameterization approach, which is only based upon averaged properties, is insufficient. Understanding the deterministic correction term in physical terms will shed light on the impact of stochastic parameterizations on systematic model errors and, hopefully, compensating model errors.

Recent findings from such rigorous derivations suggest that when the time scales of the processes we need to parameterize are not very different from those of the explicitly resolved dynamics – if we are in a grey zone - memory terms can become important. This is especially relevant for developing scale-aware parameterizations, where it is difficult to control the time scale separation as the spatial resolution is altered.

Of course, the stochastic approach is not a panacea for the subgrid-scale parameterization problem and persistent model biases. Stochastic approaches must complement developments in the deterministic physical process parameterizations and dynamical core, as motivated e.g. by Stevens and Bony (2013) and Jakob (2014). Nevertheless, it is our conviction, that basing stochastic parameterizations on sound mathematical and statistical physics concepts will lead to substantial improvements in our understanding of the Earth system as well as increased predictive capability in next generation weather and climate models.

**Acknowledgements**

The idea for this article was conceived at the workshop on "Stochastic Parameterisation in Weather and Climate Models" held at the Meteorological Institute, University of Bonn, Germany, September 16-19, 2013. We thank the Meteorological Institute for hosting the workshop and the VolkswagenStiftung for financial support (Ref: 86951). We thank Dr. Karsten Peters and two anonymous reviewers as well as the editor Dr. Brian Etherton for insightful and detailed suggestions, which greatly improved the manuscript. Thanks also to Drs. Hugh Morrison and Wojciech Grabowski for thoughtful comments on an earlier version of the manuscript.

# LIST OF FIGURES

# FIGURES

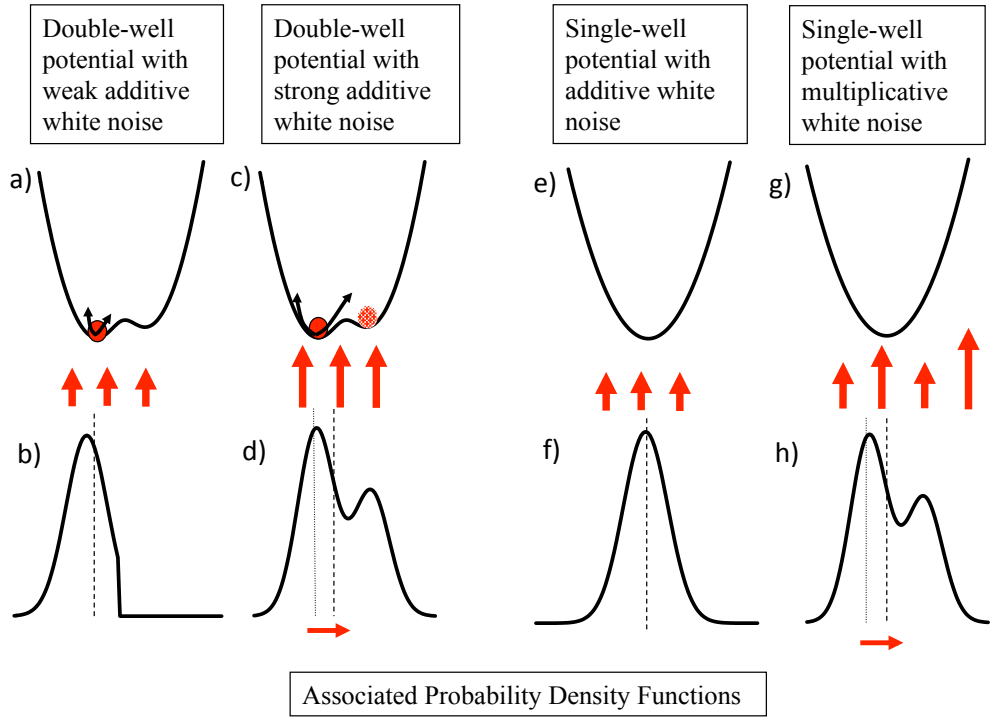

Figure 1: System characterized by a,c) double-potential or e,g) single-potential well and their associated probability density functions (PDFs). If the noise is sufficiently small (a) and under appropriate initial conditions, the system will stay in the deeper potential well and the associated probability density function of states will have a single maximum (b). As the amplitude of the noise increases, the system can undergo a noise-transition and reach the secondary minimum in the potential (c) leading to a shifted mean and increased variance in the associated probability density function (d). A linear system characterized by a single potential well and forced by additive white noise (e) will have a unimodal PDF. However, when forced by mutliplicative (state-dependent) white noise (g), the noise-induced changes the single-well potential of the unforced system, so that the effective potential including the effects of the multiplicative noise has multiple wells and the associated PDF becomes bimodal (h).

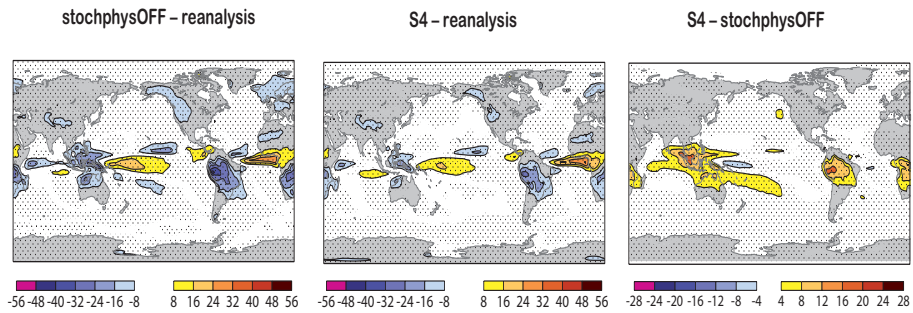

Figure 2: Bias in the top of the atmosphere net longwave radiation (outgoing longwave radiation) in W m−2 in DJF for the period 1981-2010. Simulations are conducted with ECMWF's seasonal forecasting System 4 with (S4) and without (*stochphysOFF)* stochastic parameterizations. Left and middle panels show difference from ERA-Interim reanalysis, right panel difference between experiments. Significant differences at the 95% confidence level based on a two-sided *t*-test are hatched. Adapted from Weisheimer et al. (2014).

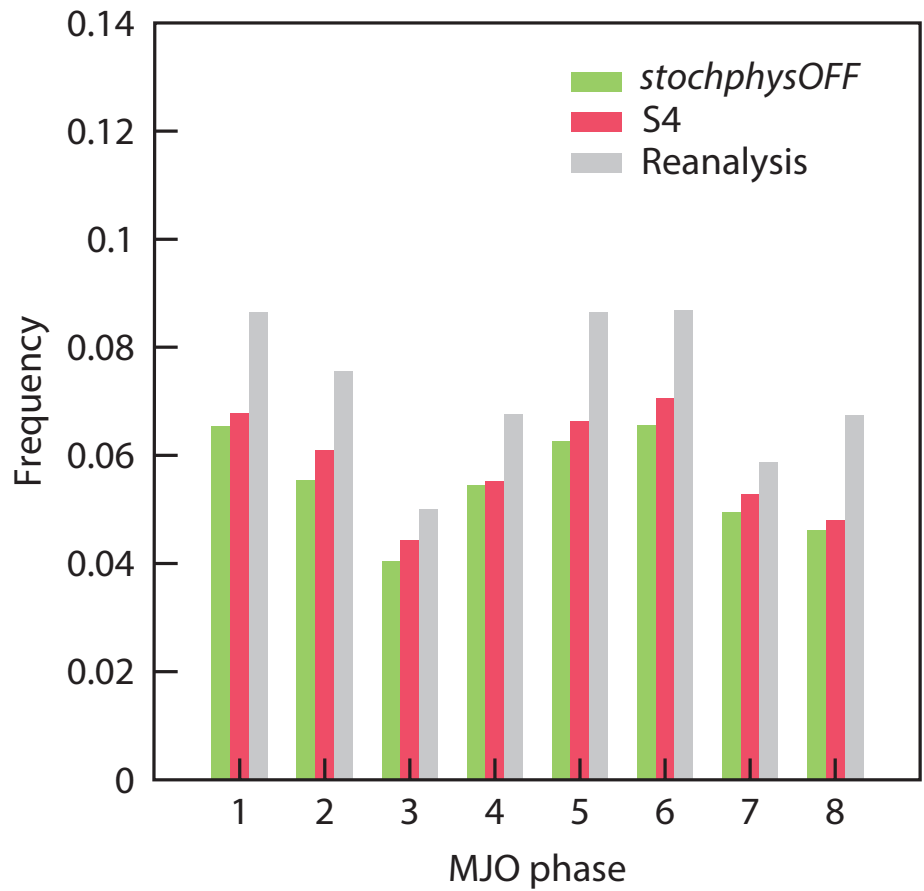

Figure 3: Relative frequencies of MJO events in each of the eight MJO phases for the period 1981-2010. Simulations are conducted with ECMWF's seasonal forecasting System 4 with (red) and without (green) stochastic parameterizations. Relative frequencies in ERA-Interim reanalysis in are shown in grey. From Weisheimer et al. (2014).

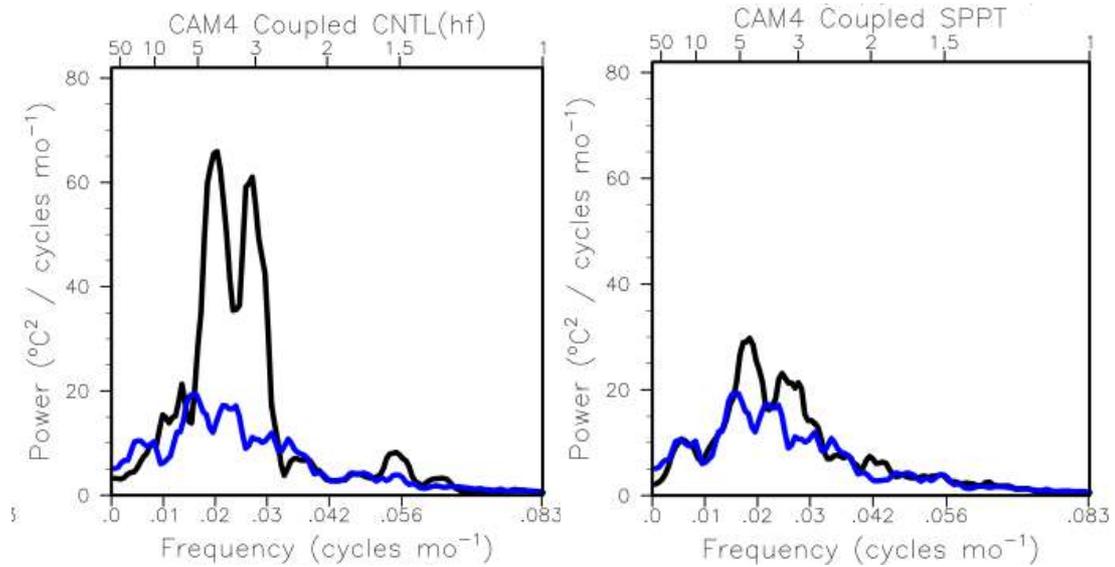

Figure 4: Power spectra of average sea surface temperature in the Nino 3.4 region in 135 year long simulations with the Community Earth System Model. Compared to HadISST observations (blue), the simulation has three times more power for oscillations with periods between 2 to 4 years (left). When the simulation is repeated with the stochastic parameterization SPPT, the temperature variability in this range is reduced, leading to a better agreement between the simulated and observed spectra (right). Adapted from Christensen et al. (2016)©American Meteorological Society. Used with permission.

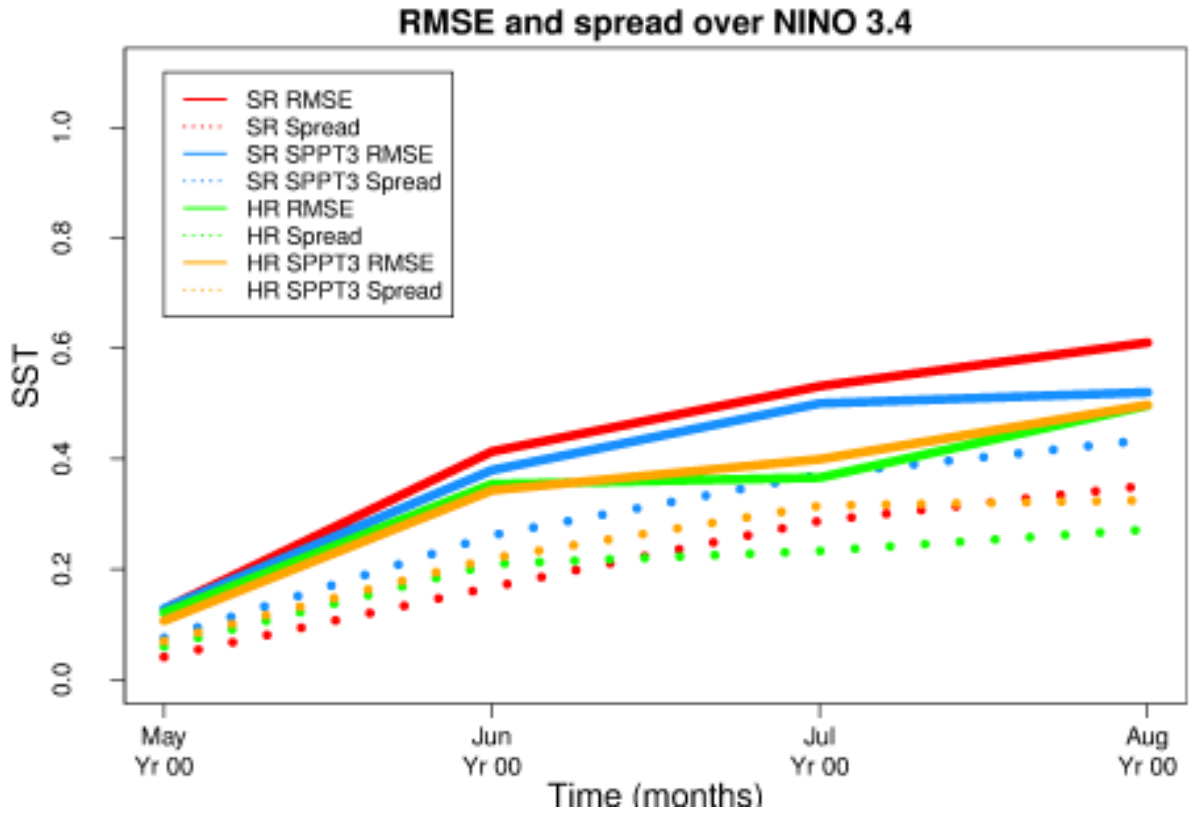

Figure 5: Niño 3.4 SST root mean square error (lines) and ensemble spread (dots) according to forecast time in EC-Earth 3 seasonal re-forecast experiments initialized in May 1993-2009 with standard (SR) or high resolution (HR) atmosphere and ocean components, with and without activating a 3-scale SPPT perturbation (SPPT3) method in the atmosphere.

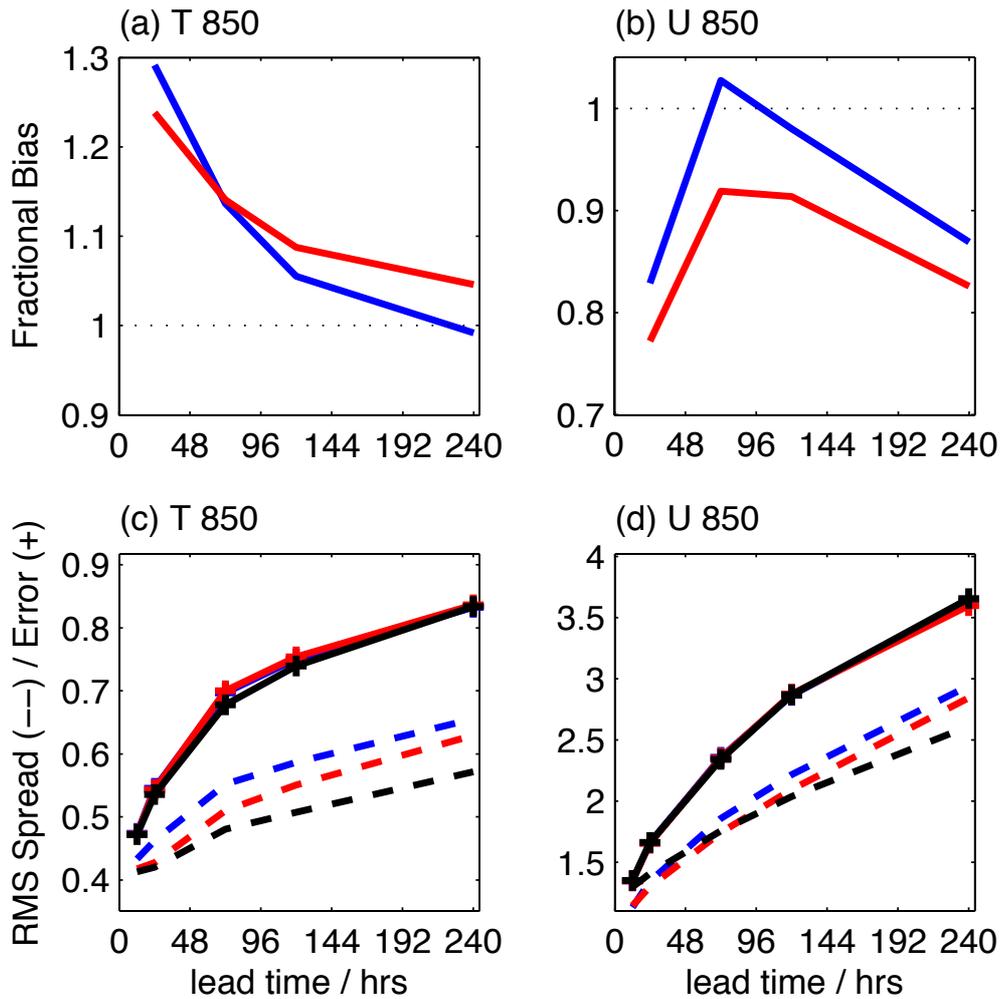

Figure 6: Forecast diagnostics as a function of time for the operational ECMWF IFS (black), fixed perturbed parameter (blue) and stochastically varying perturbed parameter (red) ensemble forecasts. Top: Forecast bias for (a) T850 and (b) U850 shown as a fraction of the bias for the operational system: BIAS /BIASoper. Bottom: Root mean square ensemble spread (dashed lines) and root mean square error (solid lines) for (c) T850 and (d) U850. Diagnostics are averaged over the region 10S-20N, 60-180E. Adapted from Christensen et al. (2015b)©American Meteorological Society. Used with permission.

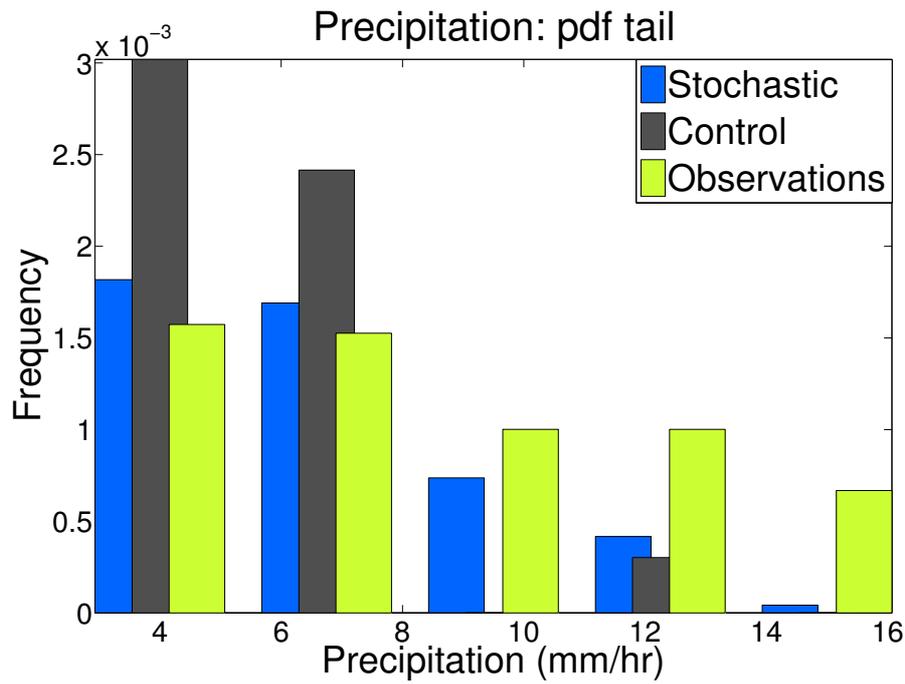

Figure 7: The right tail of the probability density function of summer season hourly precipitation from a 50-member ensemble of one year single column model simulations with stochastic (blue) and conventional parameterizations (black) of land cover over a model grid box encompassing the US Department of Energy's Atmospheric Radiation Measurement program's site in Lamont, Oklahoma. Observations are shown in green. The large-scale forcing for the single column model simulations are generated from a present day CESM simulation at a spatial resolution of about 2.8°x2.8°. Adapted from Langan et al. (2014) ©Elsevier. Used with permission.

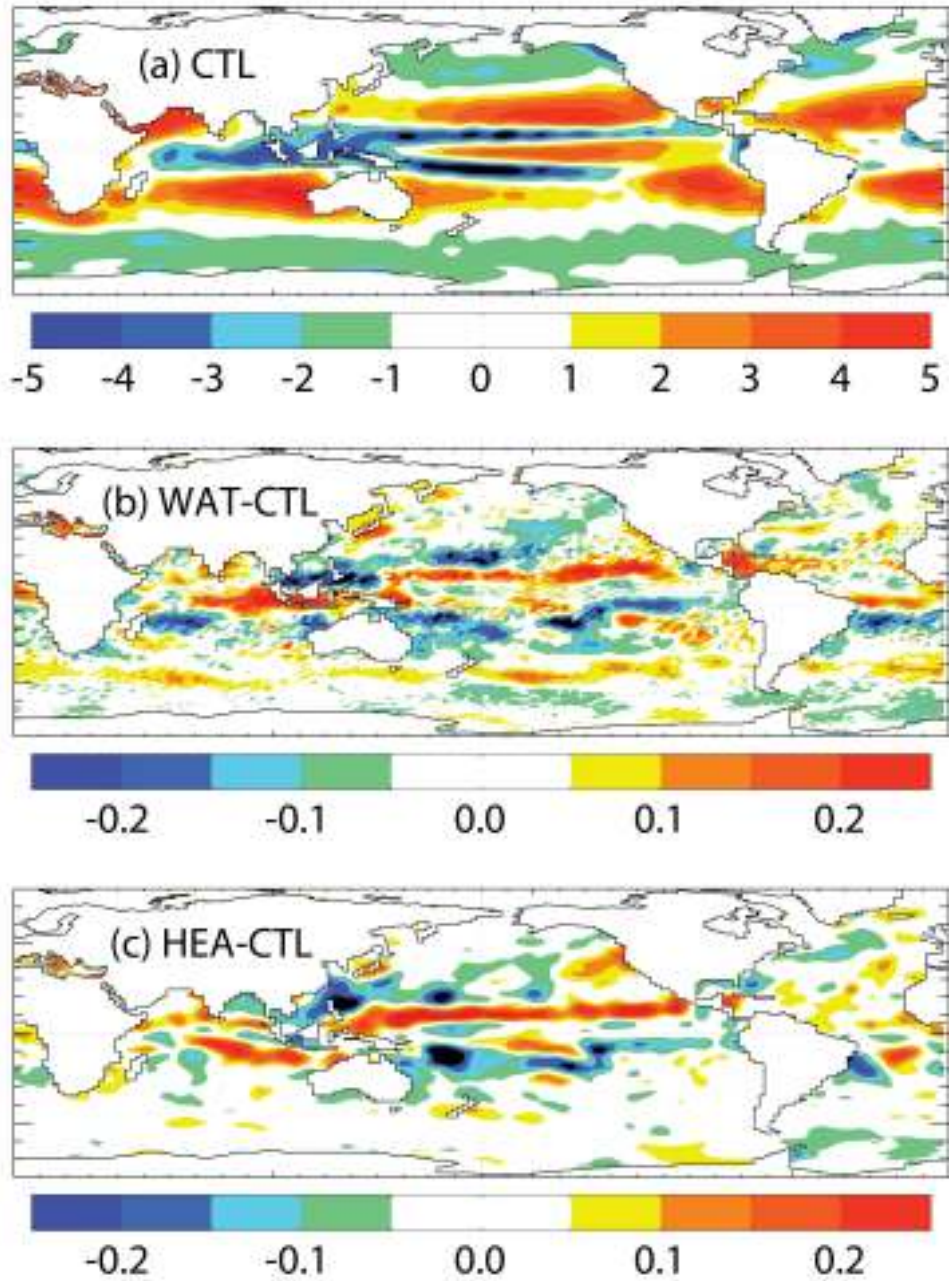

Figure 8: Maps of the century-mean net upward water flux (mm/day) at the sea surface in (a) a control integration of a coupled climate model. (b) Difference from the control for an experiment in which the net fresh water flux across the air–sea interface is stochastically perturbed before being passsed to the ocean. c) Difference from the control for an experiment in which the net heat flux across the air–sea interface is stochastically perturbed before being passsed to the ocean. From Williams (2012)©American Geophysical Union. Used with permission.

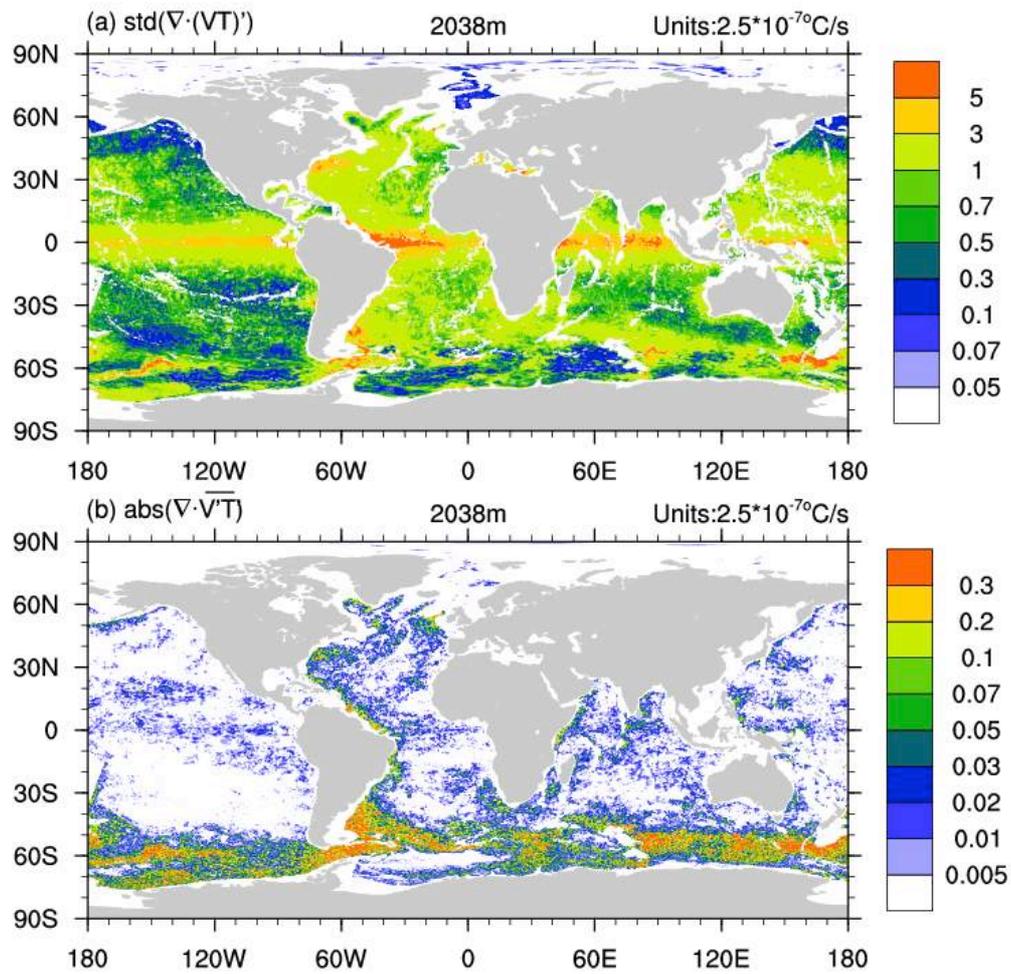

Figure 9: Top: Amplitude of fluctuations of the eddy forcing as measured by the standard deviation of divergence of eddy flux in a 1/10 degree OGCM. Bottom: Mean eddy forcing measured by the magnitude of the mean divergence of eddy heat flux in the same OGCM. The amplitude of the fluctuations is about one order of magnitude larger than the mean eddy forcing. Adapted from Li and von Storch (2013)©American Meteorological Society. Used with permission.

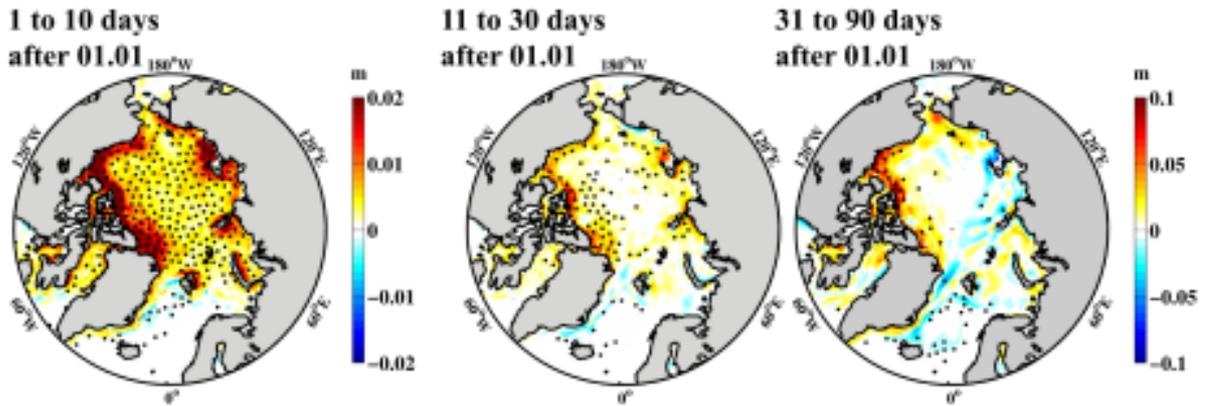

Figure 10: Difference in mean standard deviation of sea ice thickness forecasts (meters) between ensembles generated by stochastic ice strength as well as atmospheric initial perturbations and ensembles generated solely by atmospheric initial perturbations, averaged for days (left) 1 to 10, (middle) 11 to 30, and (right) 31 to 90 after initialization at 00 UTC on 1 January. Stippled areas indicate differences statistically significant at the 5% level, using a two-tailed $F$ test. Note the different contour intervals. Adapted from Juricke et al. (2014)©American Geophysical Union. Used with permission.

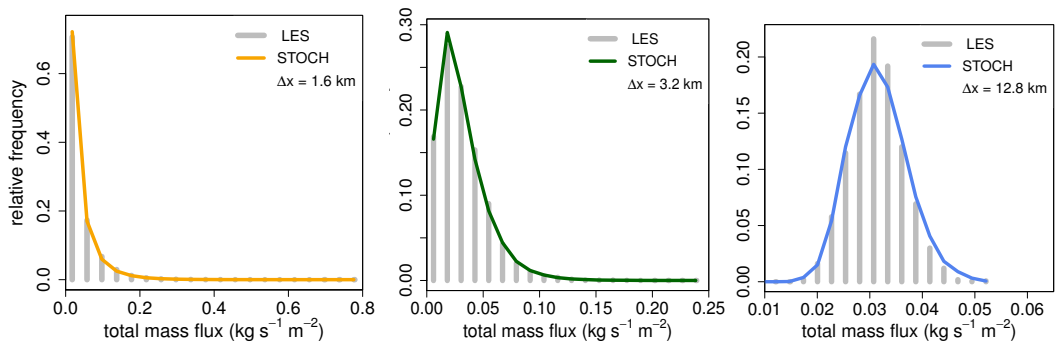

Figure 11: Histograms of the subgrid cloud-base mass flux, resulting from the stochastic shallow cumulus cloud scheme (STOCH) and coarse-grained large-eddy simulation (LES), are compared for three horizontal grid resolutions of 1.6 km, 3.2 km and 12.8 km. Adapted from Sakradzija et al. (2015).

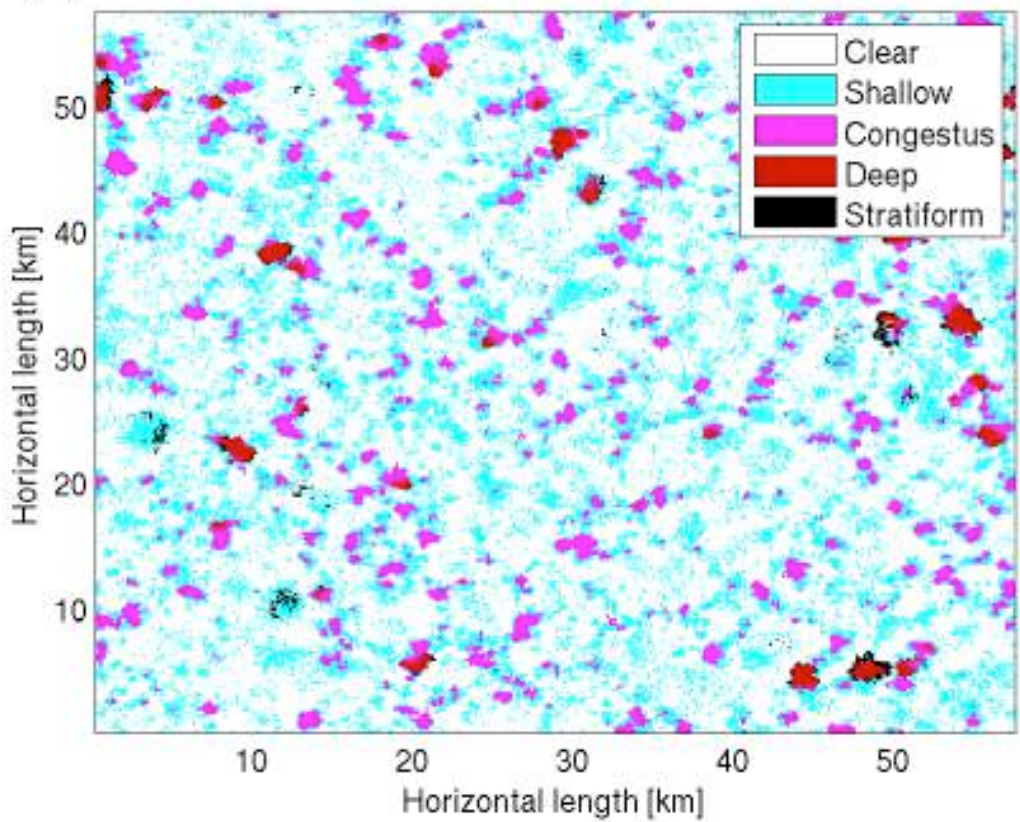

Figure 12: Snapshot of the spatial field of convective states obtained from Large Eddy Simulation data. The distinction between the various convective states was based on cloud top height and rainwater content. Adapted from Dorrestijn et al. (2013a).

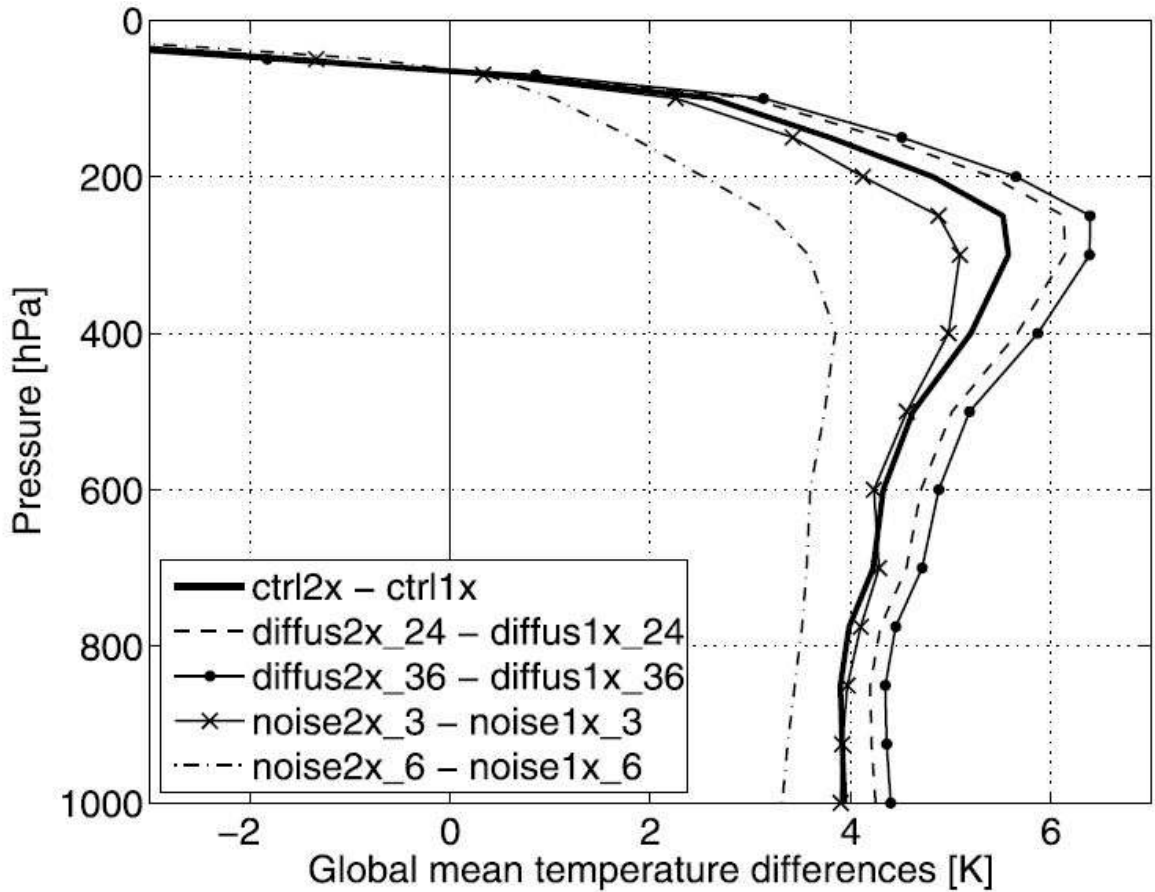

Figure 13: Climate responses of global mean temperature to a CO2 doubling (2x CO2 minus 1x CO2) obtained from the ECHAM5/MPIOM-experiments with different representations of small-scale fluctuations: 'diffus' refers to experiments in which the strength of horizontal diffusion is varied; 'noise' refers to experiments in which white noise is added to small scales of the atmospheric model ECHAM5. From Seiffert and von Storch (2008)©American Geophysical Union. Used with permission.

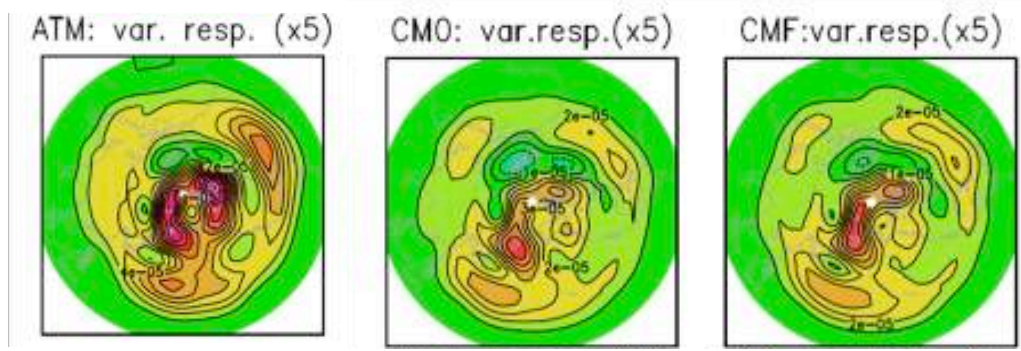

Figure 14: (Left) The response in mean streamfunction variance of a barotropic-vorticity-equation to an anomalous vorticity forcing at latitude 45N and longitude 210E projected onto 90 EOFs (left). The simulation of this response by a (middle) 90-EOF climate model with unmodified subgrid-scale parameterization (relative error 0.527), and by a (right) climate model with subgrid-scale parameterization corrected by FDT (relative error 0.342). Adapted from Achatz et al. (2014)©American Meteorological Society. Used with permission.

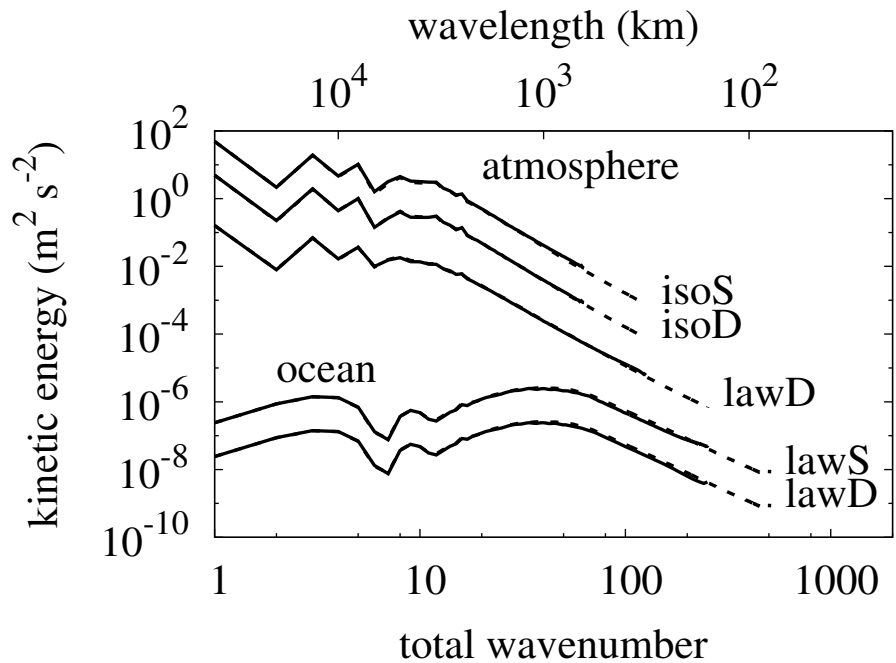

Figure 15: Top: Comparison of the upper level kinetic energy spectra of a two level benchmark simulation (dashed line) with associated LES (solid line) at various resolutions for: atmospheric isotropic stochastic (isoS) LES (top spectra); atmospheric isotropic deterministic (isoD) LES (second spectra); atmospheric deterministic scaling law (lawD) LES (third spectra); oceanic stochastic scaling law (lawS) LES (forth spectra); and oceanic deterministic scaling law LES (bottoms spectra). Top spectra has the correct kinetic energy, with the others shifted down for clarity.